\documentclass[conference, 10pt]{IEEEtran}

\newcommand{\isrevision}{1}
\usepackage{graphicx}
\usepackage{caption}
\usepackage{subcaption}
\usepackage{url}
\graphicspath{ {./images/} }
\usepackage{amsthm}
\usepackage{amsmath} 
\usepackage{xspace}
\usepackage{multicol}
\usepackage[T1]{fontenc}
\usepackage[utf8]{inputenc}
\usepackage{authblk}
\usepackage{comment}
\usepackage{lipsum}
\usepackage{mathtools}
\usepackage{cuted}
\usepackage{authblk}
\usepackage{soul}
\usepackage[normalem]{ulem}

\makeatletter
\renewcommand\AB@affilsepx{, \protect\Affilfont}
\makeatother

\usepackage{tabularx}

\usepackage[normalem]{ulem} 

\makeatletter
\def\url@allbreakstyle{%
  \def\UrlBreaks{\do\.\do\@\do\\\do\/\do\!\do\_\do\|\do\;\do\>\do\]%
    \do\)\do\,\do\?\do\'\do+\do\=\do\#%
    \do A\do B\do C\do D\do E\do F\do G\do H\do I\do J\do K\do L\do M%
    \do N\do O\do P\do Q\do R\do S\do T\do U\do V\do W\do X\do Y\do Z%
    \do a\do b\do c\do d\do e\do f\do g\do h\do i\do j\do k\do l\do m%
    \do n\do o\do p\do q\do r\do s\do t\do u\do v\do w\do x\do y\do z%
    \do 0\do 1\do 2\do 3\do 4\do 5\do 6\do 7\do 8\do 9%
  }%
}
\def\url@restrictedbreakstyle{%
  \def\UrlBreaks{\do\.\do\@\do\\\do\/\do\!\do\_\do\|\do\;\do\>\do\]%
    \do\)\do\,\do\?\do\'\do+\do\=\do\#}%
}
\makeatother

\ifnum\isrevision=1
    \usepackage[textsize=scriptsize]{todonotes}
    \newcommand{\stkout}[1]{\ifmmode\text{\sout{\ensuremath{#1}}}\else\sout{#1}\fi}
    
\else
    \usepackage[disable]{todonotes}
    
\fi

\usepackage{hyperref}

\title{Can Vehicular Cloud Replace Edge Computing?}

\author{
Rosario Patanè \\ University Paris-Saclay, France \\ \emph{rosario.patane@universite-paris-saclay.fr}
\\ \\
Andrea Araldo\\ SAMOVAR, Télécom SudParis, IPP, France \\ \emph{andrea.araldo@telecom-sudparis.eu}
\and Nadjib Achir\\Universit\'{e} Sorbonne Paris Nord/INRIA, France \\ \emph{nadjib.achir@inria.fr}
\\ \\
Lila Boukhatem \\University Paris-Saclay, LISN, CNRS, France \\ \emph{lila.boukhatem@universite-paris-saclay.fr}
}

\begin{document}

\maketitle
\begin{abstract}
Edge computing (EC) consists of deploying computation resources close to the users, thus enabling low-latency applications, such as augmented reality and online gaming. However, large-scale deployment of edge nodes can be highly impractical and expensive.
Besides EC, there is a rising concept known as Vehicular Cloud Computing (VCC). VCC is a computing paradigm that amplifies the capabilities of vehicles by exploiting part of their computational resources, enabling them to participate in services similar to those provided by the EC. The advantage of VCC is that it can opportunistically exploit part of the computation resources already present on vehicles, thus relieving a network operator from the deployment and maintenance cost of EC nodes.
However, it is still unknown under which circumstances VCC can enable low-latency applications without EC. In this work, we show that VCC has the potential to effectively supplant EC in urban areas, especially given the higher density of vehicles in such environments.
The goal of this paper is to analyze, via simulation, the key parameters determining the conditions under which this substitution of EC by VCC is feasible. In addition, we provide a high level cost analysis to show that VCC is much less costly for a network operator than adopting EC.
\end{abstract}


\IEEEpeerreviewmaketitle

\section{Introduction}
\label{sec:introduction}
Following the emergence of Cloud Computing (CC), it became apparent that despite the abundance of resources such as processors, memory, and storage, certain applications with time-sensitive requirements still face significant latency issues~\cite{Tang2022}. In response, a new computational paradigm known as Edge Computing (EC) emerged. Unlike CC, EC brings computing resources geographically closer to the end user. Such resources can be deployed in Base Stations (BS) or Road-Side Units (RSUs), for example, leading to a substantial latency reduction. It is generally accepted that CC resources can be assumed to be infinite, while EC resources are assumed limited~\cite{Liu2021}. 

In addition to CC and EC, interest in Vehicular Cloud Computing (VCC) has grown recently. Nowadays, vehicles are equipped with the most efficient computational resources, such as CPUs and GPUs~\cite{teslaGPU}. Although these resources are primarily designed for tasks related to vehicle perception and navigation, with latency requirements of around 30ms for higher levels of automation ~\cite{SEArequirements}, it is reasonable to consider that their usage for such tasks would not consume their full capacity~\cite{Zhang2015}. Therefore, we can opportunistically exploit the unused resources to compute tasks offloaded from other user devices, such as smartphones, in the vicinity of the vehicles. It's important to emphasize that the vehicular resources are already in place; hence, the cost of utilizing the VCC is lower in comparison to that of EC nodes  deployment~\cite{Hou2016}.

Many research works have introduced a three-tier architecture that integrates the computing capabilities of EC, CC, and VCC. In most of these studies, VCC has commonly been positioned as a contingency solution~\cite {Liu2021, Zhang2022,Guo2019} in which VCC is used only if needed. 

In this work, we present arguments demonstrating that VCC can replace EC under specific conditions. We adopt a radically different point of view by examining VCC as a potential \emph{replacement for} EC. Indeed, VCC offers the means to take over from EC in certain areas, allowing the execution of low-latency applications in close proximity to end users. Furthermore, leveraging VCC eliminates the requirement for an expensive and extensive deployment of EC and highlights the potential of VCC in enabling a smooth shift toward a long-term EC deployment. 
This paper's main contribution demonstrates how VCC can be a cost-effective alternative to EC by capitalizing on the abundant computational resources available in VCC~\cite{teslaGPU}. This means that there is no additional cost for the network operator in resource deployment. We analyze and present the key parameters determining the circumstances under which this substitution is effective. These parameters encompass the workload of the task, the number of end-users, the number of vehicles involved in the system, and the amount of computational resources per vehicle.

The remainder of this paper is organized as follows. In Sec.~\ref{sec:related_work}, we expose the related work and describe the system model in Sec.~\ref{sec:system_model}.
We detail the performance evaluation and simulation results in Sec.~\ref{sec:numerical_results}. Finally, Sec.~\ref{sec:conclusion} concludes the paper.

\section{Related work}
\label{sec:related_work}

Task offloading is a technique for transferring tasks from a mobile device to a more powerful remote server for processing purposes. This technique is useful for improving the performance and extending the battery life of mobile devices, but it raises challenges related to latency, cost, security, and network availability~\cite{Feng2022}. In~\cite{OffloadTask,Mazouzi2019}
, the authors presented effective methods for the selection of tasks to be offloaded. A task is offloaded if some criteria have been met, for example, if the tasks have no inter-dependency or device dependency. In our study, we operate under the assumption that the decision regarding task offloading has already been made, and our primary focus is solely on tasks that are eligible for offloading.


In~\cite{Chen2022}, the authors state that 
in peak workload periods, the tasks' queuing waiting time in EC increases significantly. 
This affects the EC's responsiveness for offloading time-sensitive application tasks under very high workloads. Hence, EC may not always the best choice for task offloading and an alternative solution is needed. In our paper, we aim to demonstrate that the EC is not the one and only possible solution for task offloading.

In~\cite{Zhang2022}, the authors propose a model comprised of three layers: VCC, EC, and CC. All of these components collaborate at the same level (no hierarchy of use is present) to provide the necessary resources to accomplish task offloading requested by wireless devices embedded in vehicles. 
In our work, we evaluate the feasibility of the substitution of EC by VCC, thus, the computation paradigms are not at the same level. 
Furthermore, mobile devices generate offloading requests, but they cannot communicate directly with vehicles. Therefore, the communication passes necessarily by an Access Point (AP), considered as the \emph{decision center}. 

In~\cite{Guo2019, Liu2022}, the authors consider two types of vehicles: \emph{customer vehicles} and \emph{servers vehicles}. When the \emph{customer vehicles} require additional resources to complete their tasks, the \emph{servers vehicles} make their resources available to help them complete their task computation. The offloading decision is made in a decentralized way and taken by each vehicle. However, in our work, we opt for a centralized approach. A Controller situated alongside the Access Point (AP) has a comprehensive overview of the available vehicles and determines the task offloading destination. The rationale behind this choice is elaborated in \S\ref{sec:offloading-architecture}. 

In~\cite{Zhang2015}, end-user devices dispatch tasks for potential offloading to either a Wi-Fi AP or a 3G/4G BS. These tasks are subsequently directed to a Cloudlet. Acting as a central hub, this edge node assesses whether a task is best suited for offloading to the CC, the Cloudlet, or the VCC. Our work adopts a similar architectural model described in \S\ref{sec:system_model}. 
However, in our work, instead of introducing yet another offloading strategy that accounts for CC, EC, and VCC collectively, we concentrate on exploring whether and under what circumstances VCC can effectively replace EC, particularly in cases where deploying EC is economically impractical.

\section{System Model}
\label{sec:system_model}

\begin{figure} [t]
    \centering
    \includegraphics[width=0.4\textwidth]{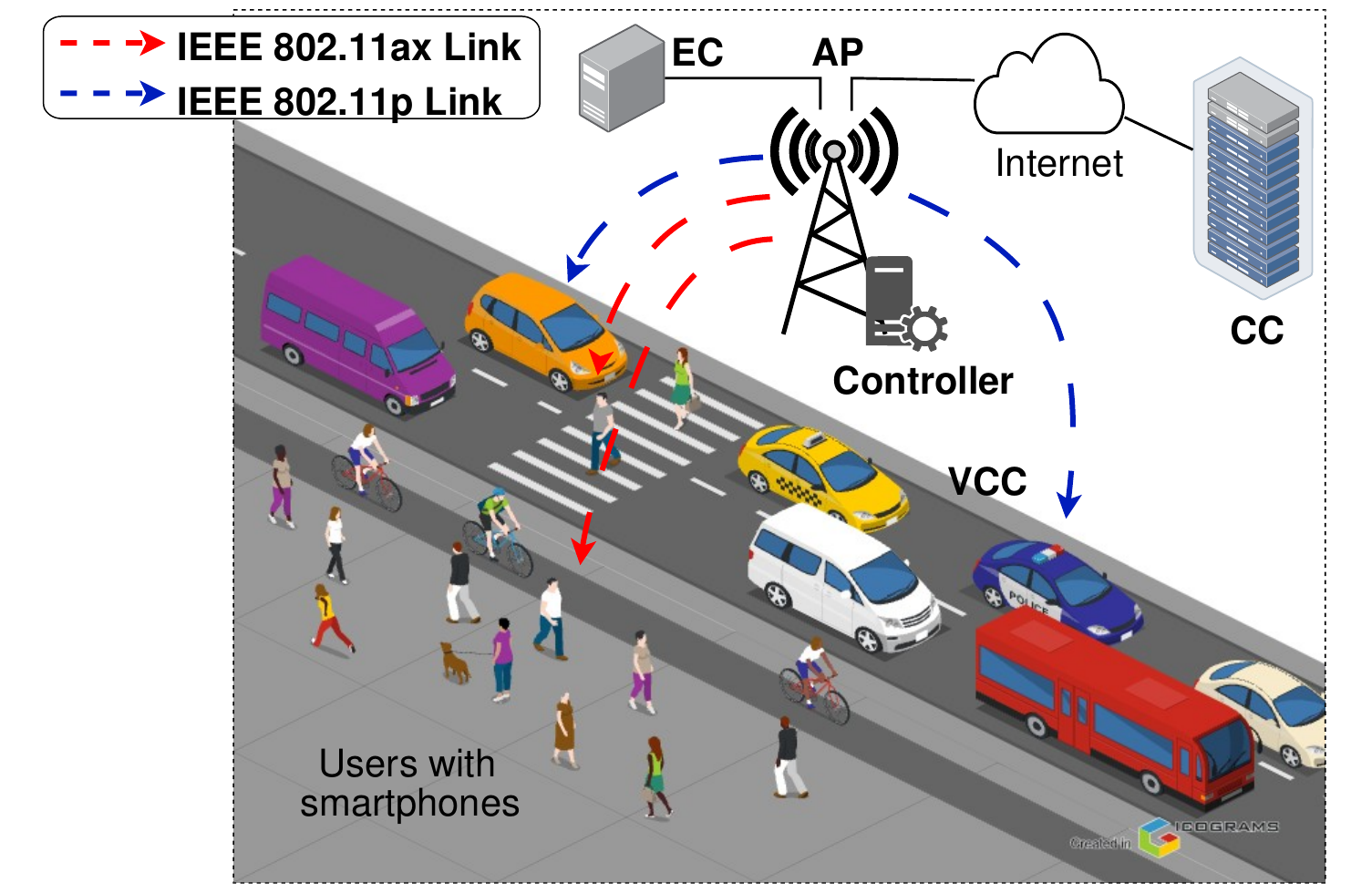}
    \caption{Model architecture}
    \label{fig:CoE}
\end{figure}

The system architecture is depicted in Fig.~\ref{fig:CoE}. End users are considered pedestrians which possess end devices such as smartphones, smartwatches, or smart glasses. While these devices may not be able to compute all the required tasks locally, they can offload some tasks to the network. The offloading is done to preserve the limited energy available in the device's battery or when the computational demands exceed the device's capacity~\cite{offloadIsBetter}.
The decision of whether to offload or not~\cite{OffloadTask} is beyond the scope of this paper. Here, we exclusively consider tasks that the user devices have already chosen to offload. 

\subsection{Offloading architecture}
\label{sec:offloading-architecture}
The end devices send task offloading requests to the Wi-Fi AP (which constitutes the RSU). 
We assume IEEE802.11ax (Wi-Fi 6) as the communication technology between the end devices and the AP. We also assume that all vehicles use the IEEE802.11p standard (Wave) for communication between the AP (the RSU) and vehicles.
Finally, we assume that the AP is connected by optical fiber to the Internet, enabling connectivity to CC through a UDP connection.
Co-located with the AP, a \emph{Controller} decides whether to offload the task to EC node, VCC node, or CC, according to an offloading strategy. We assume that the network operator directly manages the EC, while the VCC resources are managed by the respective car manufacturers. It is assumed that the necessary agreements, protocols, and APIs are in place to facilitate task offloading.
Finally, for simplicity, we consider a single restricted area with only one AP in this study. 
Larger APs and RSUs deployment areas will be considered in our future works.
In this work, we assume that the user device initiates the task offloading request and sends it to the AP. Via periodic vehicle-to-AP beaconing (\S\ref{sec:scenarios}), the Controller has a complete view of the VCC and maitains a list all the available vehicles within its coverage. 
We assume that issues related to privacy, energy consumption, and incentives for car owners are addressed as described in the literature~\cite{Liu2022, Wei2022}. Such issues are outside of the scope of this paper.

\subsection{Considered offloading scenarios and network protocols}
\label{sec:scenarios}
We define the following offloading scenarios:

\begin{itemize}
    \item \emph{ECFirst}: Here, the scenario includes only EC and CC for task offloading. To minimize latency, the Controller prioritizes offloading tasks to the EC unless the EC resources are fully utilized, i.e. the waiting queue at the EC is full. In this case, the Controller offloads the task to the CC. This strategy ensures low latency by leveraging as many EC capabilities as possible while seamlessly using CC resources as a fallback option when necessary.
    \item \emph{VCCFirst}: Here, the strategy includes only VCC and CC, suppressing the need for EC. When the Controller receives a task offloading request, it checks for the availability of vehicles in the coverage of the AP. If vehicles are present, the task is randomly assigned to one of the available vehicles. However, the task is offloaded to the CC if no vehicles are available or all the resources in vehicles are exhausted. 
\end{itemize}

As we can see, we have explicitly omitted considering a strategy composed of VCC and EC. Indeed, using a Vehicular-Edge Computing strategy is not indicative since it behaves similarly to VCCFirst with the EC instead of the CC. However, to prove that the substitution of EC by VCC is feasible, we study the standalone paradigm (VCC, EC) with the CC as a backup. 

In the \emph{ECFirst} scenario, we consider a node with EC resources equipped with a CPU, colocated with the AP. The Controller can continuously monitor the resource utilization in this node, and if the resources become exhausted, it sends the requests to the CC.
In the \emph{VCCFirst} scenario, we assume that each vehicle periodically transmits a beacon to the Controller at a frequency of $f$ (in the simulations, we set $f= 10Hz$).
These beacons serve as communication messages from the vehicles to the Controller, providing essential information about their current status such as their current amount of available resources. 

The beacons can be seen as an extension of the Cooperative Awareness Messages (CAM) transmitted by each vehicle according to the Wave standard~\cite{WaveStandard}. By receiving and processing these beacons, the Controller can maintain an updated list of currently available vehicles under its coverage and capable of hosting tasks. If the controller has not received a beacon from a vehicle within a given period (set to $500ms$ in simulations), it removes that vehicle from the list. 
As we can see, the AP/RSU plays a dual role as it bridges different network protocols and incorporates a Wave communication device implementing the IEEE802.11p physical standard. Furthermore, the adoption of this technology is justified by its present readiness and availability compared to the still-in-progress deployment of 5G-V2X technology~\cite{Waveisbetter}.

\subsection{Task model}
\label{sec:task-model}
We model task $i$ as a tuple $(W^i, S^i, R^i)$, where workload $W^i$ (measured in Million Instructions (MI)) is the amount of instructions required to execute the task,  $S^i$ is the input task size (in KBytes) and $R^i$ is the amount of data to transmit back to the end user after the computation.
The offloading process is executed as follows. Once the end-user device decides to offload a task, it sends it and the corresponding information to the AP. The Controller chooses where to offload the task (EC, VCC, or CC) depending on the strategy which is then queued and executed. If CC or EC process the task, they send the result directly to the end-user. In case the task is executed at a vehicle, the result is first sent to the AP which then relays it to the end-device in compliance to the technologies incompatibility. 

\subsection{Cloud model}
\label{sec:cloud}
We assume that CC has an infinite resource capacity. Any task offloaded to the CC is immediately processed without any waiting time.

The offloading time $T^i_\text{CC}$ of task $i$ to CC corresponds to the time between the moment the end device of the respective end-user sends a task offloading request and the moment the end device receives the offloading result. It can be divided into several components, all expressed in seconds:
\begin{itemize}
    \item The \textit{Uplink time}, noted $T^i_\text{up, AP}$, which is the time required for the task to reach the AP from the end device.
    \item The \textit{Uplink Core Network (CN) time}, noted $T^i_\text{up, CN}$, represents the time to pass through the edge node, the CN/Internet, and reach the cloud node.
    \item The \textit{Elaboration time} $T^i_\text{elab}$  is the time needed for the task computation. We have $T^i_\text{elab}=W^i/C_\text{CC}$, where $C_\text{CC}$ is the computational capacity of the cloud, expressed in Million Instructions Per Second (MIPS). 
    \item The \textit{Downlink time}, which is the time required for the result to be sent back to the user who requested the task offloading. It comprises $T^i_\text{down, AP}$ and $T^i_\text{down, CN}$. 
\end{itemize}

The total task offloading time (a.k.a. task response time) to CC, noted $T^i_\text{offloading, \text{CC}}$, can then be expressed as follow: 
\begin{multline}
\label{eq:cloud_offloading_time_1}
        T^i_\text{offloading, \text{CC}}= \\T^i_\text{up, \text{AP}}+ T^i_\text{up, \text{CN}}+T^i_\text{elab}+T^i_\text{down, \text{AP}}+T^i_\text{down, \text{CN}}= \\T^i_\text{up, \text{AP}}+ T^i_\text{up, \text{CN}}+\frac{W^i}{C_\text{CC}}+T^i_\text{down, \text{AP}}+T^i_\text{down, \text{CN}}.
\end{multline}

\subsection{Edge model}

The Edge node is deployed on a Wi-Fi AP. The EC framework comprises a single computational resource, such as a CPU or GPU of $C_\text{EC}$ computational capacity measured in MIPS. Consequently, to handle the temporary unavailability of resources, the EC incorporates a FIFO policy queue mechanism for storing offloading requests.

The task offloading time to the EC node, noted $T^i_\text{offloading, \text{EC}}$, can be expressed as follows: 
\begin{multline}
        T^i_\text{offloading, \text{EC}}= T^i_\text{up, \text{AP}}+T^i_\text{queue}+T^i_\text{elab}+T^i_\text{down, \text{AP}}=\\ 
        T^i_\text{up, \text{AP}}+T^i_\text{queue}+\frac{W^i}{C_\text{EC}}+T^i_\text{down, \text{AP}}
\end{multline}

\noindent where $T^i_\text{queue}$ is the task's waiting time at the EC node before being executed.

\subsection{Vehicular cloud model}
\label{sec:vehicular-cloud-model}

We denote with $T^i_\text{VCC}$ the offloading time of task $i$ on the vehicular cloud and $C_\text{VCC}$ the computational capacity of a single vehicle (we assume for simplicity all vehicles have the same capacity).
The offloading time of task $i$ to a vehicle is: 
\begin{multline}
        T^i_\text{offloading, \text{VCC}}=\\ T^i_\text{up, \text{AP}}+ T^i_\text{up, \text{Wave}}+T^i_\text{elab}+T^i_\text{down, \text{Wave}}+T^i_\text{down, \text{AP}}= \\T^i_\text{up, \text{AP}}+ T^i_\text{up, \text{Wave}}+\frac{W^i}{C_\text{VCC}}+T^i_\text{down, \text{Wave}}+T^i_\text{down, \text{AP}}
\end{multline}

\noindent where the $T^i_\text{up, \text{Wave}}$ is the offloading request transmission time on the IEEE802.11p wireless link between the AP node
and the vehicle. Similarly, $T^i_\text{down, \text{Wave}}$ is the result's transmission time on the IEEE802.11p. link from the vehicle to the AP.

\section{Simulation results}
\label{sec:numerical_results}

This section demonstrates the potential of VCC to replace EC in low-latency applications. Through extensive simulations, we pinpoint specific conditions related to vehicle speed, workload, and the number of vehicles involved, which are pivotal for this substitution. Our analysis adopts a conservative approach, subjecting VCC to "pessimistic" scenarios by considering an 802.11-based communication technology. This approach is pessimistic for several reasons. We direct all task offloading requests to pass through the AP before reaching the vehicles (\S\ref{sec:system_model}). If we were to consider 5G-V2X technology instead, tasks could be directly sent from users to vehicles. Finally, intelligent strategies could potentially enhance offloading performance on the VCC, such as selecting vehicles based on the stability of their connection to the AP. However, our primary focus is not on introducing yet another intelligent strategy for VCC but rather on determining if VCC can replace EC even under pessimistic conditions. We will explore other technologies (such as 5G) in a future work.

\subsection{Simulation environment}
\label{sec:simulation-environment}
\begin{table}
    \fontsize{8pt}{8pt}\selectfont
    \begin{tabularx}{0.48\textwidth} { 
    | >{\raggedright\arraybackslash}X 
    | >{\centering\arraybackslash}X 
    | >{\raggedleft\arraybackslash}X | } 
    \hline
    \textbf{Parameter} & \textbf{Value} \\
    \hline
    \hline
    Cloud nodes & $1$  \\
    \hline
    Number of edge nodes & $1$ (in the ECFirst scenario) or $0$ (in the VCCFirst scenario)  \\
    \hline
    Number of vehicles & Up to $50$ \\ 
    \hline
    Number of end users  & $8$~\cite{Spiceworks}  \\ 
    \hline
    Simulation duration & $120$ seconds  \\
    \hline
    Cloud computation resources & $C_\text{CC} =2356230$ MIPS~\cite{Chiappetta2020}, $\infty$ \  processors\\
    \hline
    Edge computation resources & $C_\text{EC} =749070$  MIPS~\cite{Chiappetta2019}, $1$ processor\\
    \hline
    VCC computation resources & $C_\text{VCC}=71120$  MIPS~\cite{InstrPerSec}, $1$ processor per vehicle\\
    \hline
    Task workload & $C_u=500$  MI~\cite{Fizza2019}\\
    \hline
    Task size & $D_u=4000$ bytes~\cite{Zhang2015}\\
    \hline
    Max queue length at EC & $100$ packets\\
    \hline
    Core network latency & $T^i_{up,\text{CN}}=T^i_{up,\text{CN}}=35$ milliseconds~\cite{Verizon}\\
    \hline
    End-user offloading request rate & A request every $200\text{ms}$~\cite{Zaidi2022} \\
    \hline
    Vehicle average speed & downtown traffic $13.1$km/h~\cite{speedParis}\\
    \hline

    \end{tabularx}
    \caption{Simulation default parameters.}
    \label{tab:params-table}   
\end{table}

The Network Simulator v3 (NS3) is used to simulate the system described in \S\ref{sec:system_model}. The default simulation parameters are listed in Table~\ref{tab:params-table}. In these simulations, only limited size tasks are considered and the default task size is 4KB. Observe, however, that each task may require many instructions to be executed (high workload), as explained in \S\ref{sec:task-model}. For example, a simple ''for loop'' can have a size of less than 1KB but a workload of 10000 Million Instructions. 

The vehicle mobility, is obtained from SUMO (mobility simulator) using {\it Manhattan scenarios}. Vehicles move into a grid of 200x200m composed of 2 longitudinal and 2 latitudinal streets, spaced by 100m. The AP is in the middle of the grid. Every road has two lanes in opposite directions. As for pedestrians, we assume that they are stationary during the offload of a task. This is a reasonable assumption since in such a short time (0.5s), even a ``fast'' pedestrian would have moved less than 1m. The processors used by default are \emph{AMD Ryzen Threadripper 3990X (64 cores)} for the CC, \emph{AMD Ryzen 9 3950X (16-core)} for the EC and finally \emph{ARM Cortex A73 (4-core)} for each vehicle. 
We have chosen processors available on the market, on the basis of their computing power which, according to the criteria in the literature on the computational power of CC, EC, VCC (embedded)~\cite{Fizza2019}. 
We compare the obtained offloading times with some reference application classes and their respective latency requirements (Table~\ref{tab:latency-classes}).

\begin{table}
\begin{scriptsize}
    \centering
    \begin{tabular}{|c|c|c|}
    \hline
        \textbf{Class name} & \textbf{Requirement} & \textbf{Example of applications}  \\
    \hline
        Extremely Low Latency (LL\textsuperscript{++}) 
        & $\le$ 16ms & 
        Augmented Reality~\cite{Ayoub2021}
        \\
    \hline
        Very Low Latency (LL\textsuperscript{+})
        & $\le$ 100ms
        & Augmented Reality~\cite{Ayoub2021}
        \\
    \hline
        Low Latency (LL)
        & $\le$ 500ms
        & Antivirus~\cite{kim2019}\\
    \hline
    \end{tabular}
\end{scriptsize}
    \caption{Classes of applications and latency requirements}
    \label{tab:latency-classes}
\end{table}

\subsection{Impact of the number of end users}

\begin{figure}
      \begin{subfigure}[b]{\linewidth}
       {\includegraphics[width=0.5\linewidth]{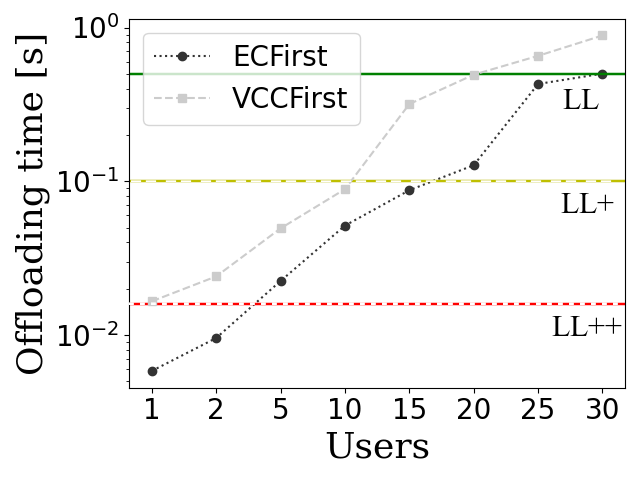}}\hfill
        {\includegraphics[width=0.5\linewidth]{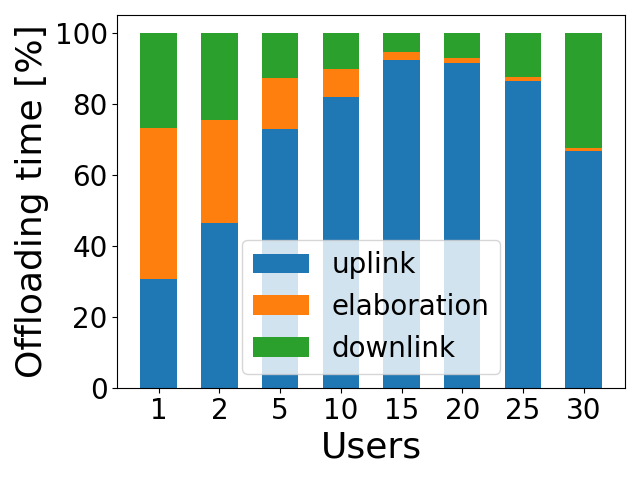}}     \end{subfigure}
      \caption{\textbf{Left}: Average offloading time ($T^i_\text{EC}, T^i_\text{VCC}$) of ECFirst and VCCFirst scenarios. \textbf{Right}: Contribution of the different components of the offloading time in the VCCFirst scenario.}
    \label{fig:off_time_random_users_cloud_edge_vcc}
\end{figure}

In Fig.~\ref{fig:off_time_random_users_cloud_edge_vcc}, the number of end devices varies. This amount is directly related to the rate of offloading task requests (as each end user generates an offloading request each $200ms$). In Fig.~\ref{fig:off_time_random_users_cloud_edge_vcc}-left, we plot three horizontal lines corresponding to the latency requirements of the considered classes of applications
(Table~\ref{tab:latency-classes}).
Fig.~\ref{fig:off_time_random_users_cloud_edge_vcc}-left shows that VCC can replace EC up to 20 end users for LL applications. However, LL\textsuperscript{++} applications cannot be supported by VCC. They strictly require EC and highly favorable conditions, such as very few users ($\le$ 3 users). The CC is only sporadically used in \textit{VCCFirst} ($\leq 1\%$).

In Fig.\ref{fig:off_time_random_users_cloud_edge_vcc}-right we observe that by increasing the number of requests, there is a growth in the offloading time because both the uplink and the downlink times are impacted. 

When the number of users increases, the impact of the elaboration time becomes negligible, and the sum of the uplink and downlink times becomes the most contributing factor to the offloading time. 
This is because as the rate of requests sent to the wireless channels increase (802.11ax and 802.11p), packets containing such requests start queuing at the Medium Access Control (MAC) layer. It is also clear that the bottleneck that prevents VCC from performing as well as EC is not computation. Therefore, installing more computation capacity in the vehicles would be useless. The bottleneck is instead the wireless network. Indeed, we observe a higher bottleneck in the communication between the vehicle and the AP compared to the observed bottleneck between the user and the AP. This is because the IEEE802.11p standard has lower throughput than IEEE802.11ax. This could be improved using 5G cellular technology.

\subsection{Impact of the number of vehicles}

\begin{figure}
     \begin{subfigure}[b]{\linewidth}
        {\includegraphics[width=0.5\linewidth]{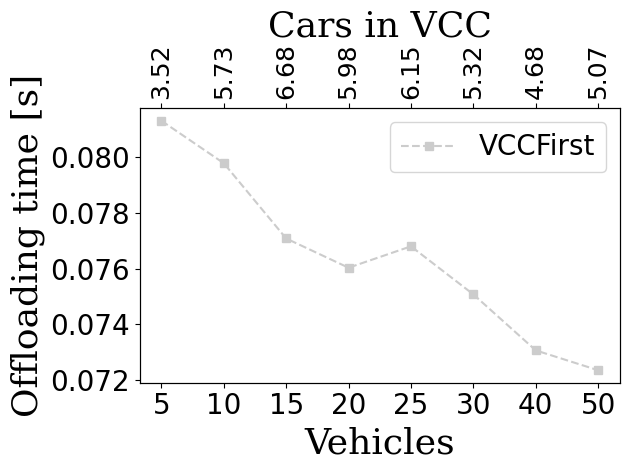}}\hfill
        {\includegraphics[width=0.5\linewidth]{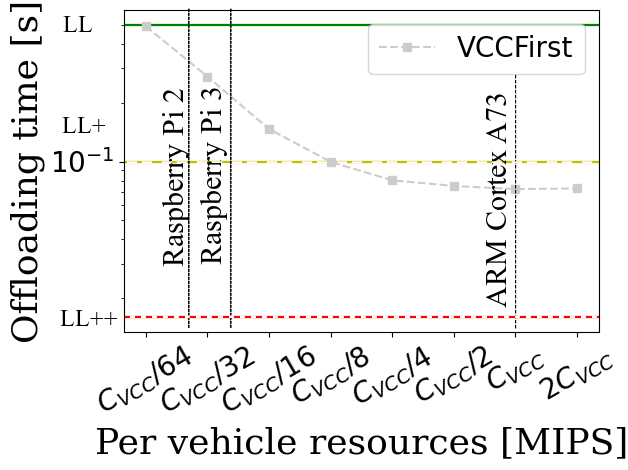}}
     \end{subfigure}
      \caption{Offloading time of VCCFirst scenario. \textbf{Left}: Impact of the number of vehicles. \textbf{Right}: Impact of computational capacity $C_\text{VCC}$ of each vehicle.}
     \label{fig:off_time_resources_in_vcc}
\end{figure}

In this scenario, we simulate the mobility of 50 vehicles in SUMO. 
In Fig.~\ref{fig:off_time_resources_in_vcc}-left, the bottom $x$-axis represents all the simulated vehicles in the scenario but only a subset of them forms the VCC. This set corresponds to the vehicles from which the Controller has received fresh beacons (\S\ref{sec:scenarios}). The top $x$-axis shows the mean number of vehicles in the VCC within the AP's coverage, averaged along the simulation time. We observe that as expected this average increases with the number of vehicles. 
In  Fig.~\ref{fig:off_time_resources_in_vcc}-left, the offloading time decreases slightly with the number of vehicles. This is because it becomes more likely to find vehicles closer to the AP and with better connectivity. This reduces the offloading time, even if we do not explicitly account for the quality of the AP-vehicle channel when selecting the vehicle onto which to offload. 
Note that, already with 5 vehicles no task needs to be offloaded to the CC (\S\ref{sec:scenarios}). 


\subsection{Impact of the computational resources deployed into vehicles}

In Fig.~\ref{fig:off_time_resources_in_vcc}-right, we consider a baseline CPU of capacity $C_\text{VCC}=71120$ MIPS. 
We then analyse the offloading time when varying the vehicles CPU capacity.
Note that for the first three values of the $x$-axis, a small part of task offloading requests (less than 1\%) had to be sent to the CC. This is because each vehicle takes a long time to process a task, which reduces the probability of finding an available vehicle. 

The overall trend, as expected, describes a diminishing offloading time due to reduced elaboration time. However, after a certain value, the increase in computational power is no more beneficial, which confirms the finding related to Fig.~\ref{fig:off_time_random_users_cloud_edge_vcc}-right: increasing computational resources into vehicles more than a certain amount is not useful.

\subsection{Impact of task workload}

\begin{figure}

       {\includegraphics[width=0.5\linewidth, height=3.65cm]{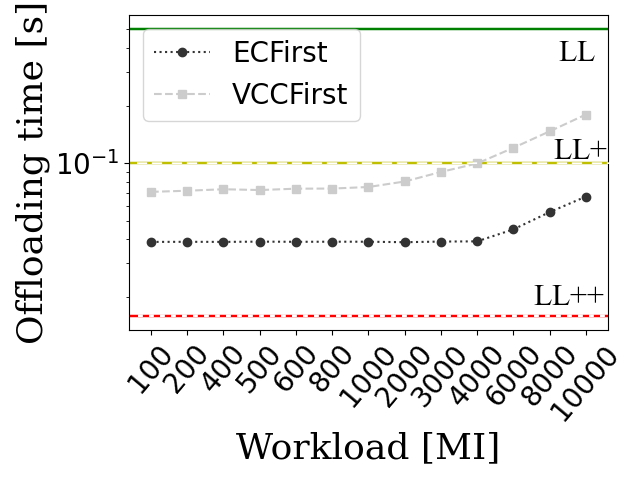}}\hfill
        {\includegraphics[width=0.5\linewidth]{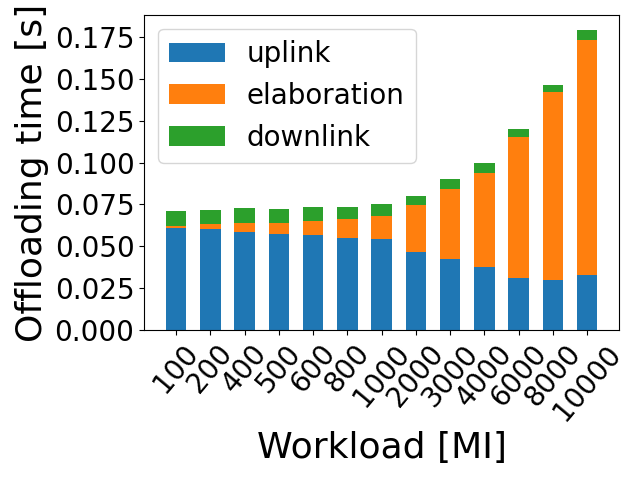}}
         \caption{\textbf{Left}: ECFirst and VCCFirst offloading time. 
         \textbf{Right}: The VCC offloading time for the \textit{VCCFirst} strategy.}
    \label{fig:off_time_workload_general}
\end{figure}

We show in Fig.~\ref{fig:off_time_workload_general} left the offloading time as a function of the task workload for each strategy. According to~\cite{Fizza2019}, the workload can be divided into three categories according to the task complexity, starting from 100 MI to 9784 MI. The task workload is 500 MI in the results in Fig.~\ref{fig:off_time_random_users_cloud_edge_vcc} and Fig.~\ref{fig:off_time_resources_in_vcc}. The application related to this workload is the~\emph{object recognition}~\cite{Fizza2019}.
We observe on the~\ref{fig:off_time_workload_general}-left figure the offloading time for the two strategies. The results show that the EC is able to serve all the possible task workloads and stay within the deadline of 100ms. This is not the case with \textit{VCCFirst}. The VCC meets the 100ms requirement until 4000 MI without the help of the CC. After 6000 MI, the VCC can meet the 500ms requirements, but this time with the help of CC at 8000-10000 MI. This means the VCC can substitute the EC for applications with task workloads only within specific ranges.

In Fig.~\ref{fig:off_time_workload_general}-right, we can observe the different components of the offloading time. We note that with the increase in the workload, the elaboration time for the VCC increases. 
The uplink time decreases because the VCC informs the EC that a  significant part of its resources is busy. This implies that the EC will send less requests to the VCC resulting in a requests failure rate of almost 10\%, for a workload of $500$MI.

\subsection{Impact of the speed}
Our results do not significantly change even if we increase the average speed of vehicles from 13.1 Km/h to 30 and 50 Km/h. This means that the time needed to process a request in a vehicle is small with respect to the movement of the vehicles and that the chances that a vehicle gets a request and then exits the coverage area before finishing are low. These results could be different when considering much higher speeds, e.g., in the highways, but such cases are not relevant for our studies, as in highway-like scenarios we do not expect anyways any end-user close to the road requesting to offload tasks. The relevant case for us is an dense urban environment, with pedestrians, which is compatible with traffic speed of about 13.1 Km/h. This is why we do not consider in our analysis any speed above 50 Km/h (which is already too high for our relevant case).

\subsection{Equipment cost analysis}
We now evaluate, at high level, the savings the network operator can achieve by adopting VCC in a certain cell instead of deploying EC.
Let us assume that the network operator pays to offload each task, with prices similar to commercial serverless computing offers~\cite{AWS}, i.e., about $2\cdot 10^{-5}$\$/req. The payment could go to the vehicle owner, or the vehicle manufacturer or an enterprise managing the VCC, it is indifferent for our analysis (note that the analysis of which should be the economic actors around VCC is outside of our scope). At the offloading rate considered here (Table~\ref{tab:params-table}), by simple calculus, the network operator would spend 1000\$ in $\sim$32 years. 1000\$ is the cost of the AMD CPU considered for EC (\S\ref{sec:simulation-environment}). Of course, the lifespan of a CPU is much less than 32 years. Moreover, if we consider the cost of the entire machine in which the CPU would be installed and, more importantly, the maintenance of each EC node over 32 years, the cost of EC in one considered cell would be even higher than 1000\$. Therefore, adopting VCC (whenever feasible) is much less costly than deploying EC for a network operator.

\section{Conclusion}
\label{sec:conclusion}
Throughout this paper, we demonstrate that edge computing can be replaced by the vehicular cloud under certain conditions. We defined a three-layered model composed by cloud, edge, and vehicular cloud. This model was useful to study the conditions under which a substitution is possible. We explored how the offloading time is impacted by varying the number of cars, users, resources per car, and workload of tasks. We concluded that for around 10 users object recognition with LL, LL\textsuperscript{+} requirements are satisfied VCC successfully replaces EC. We have observed also the impact of the uplink, computation and downlink times on the offloading time varying users and workload. Furthermore, the offloading time appears to be decreasing as the resources in the VCC increase (number of vehicles and computation capacity) and VCC can satisfy LL, LL\textsuperscript{+} task requirements. The results show also that from low to high computation intensive tasks the VCC can substitute the EC. An analysis of the impact of the speed towards the failure rate of tasks is provided. Finally, we provided an equipment cost analysis that the VCC is economically more convenient than the EC.  In our future work, we will investigate and propose intelligent strategies to improve performance parameters such as failure rate, offloading time, and energy consumption. We will also explore larger-scale edge node deployments in urban and highway scenarios based on real data traces.

\section{Acknowledgements}
This work was supported by  \href{https://digicosme.cnrs.fr/}{Labex DigiCosme, France}.

\end{document}